\documentclass[aps,pre,twocolumn,showpacs,groupedaddress,
               showkeys,amsmath,amssymb]{revtex4}
\usepackage{graphicx}
\usepackage{bm}
\usepackage{bbm}
\usepackage{epic}
\usepackage{eepic}
\usepackage{pifont}
\usepackage{nicefrac}
\begin{document}
\title{\bf A plastic flow theory for amorphous materials}
\author{V.I. Marchenko, Chaouqi Misbah}
\affiliation{P.L. Kapitza Institute for Physical Problems, RAS,
119334, Kosugina 2, Moscow, Russia\\
Lab. de Spectrom\'etrie Physique, Universit{\'e} Joseph Fourier
(CNRS) Grenoble I, B.P.\,87, Saint-Martin d'H{\`e}res, 38402 Cedex,
France} \email[]{cmisbah@ujf-grenoble.fr}
\date{\today}
\begin{abstract}

Starting from known kinematic picture for plasticity, we derive a
set of dynamical equations describing plastic flow in a Lagrangian
formulation. Our derivation is a natural and a straightforward
extension of simple fluids, elastic and viscous solids theories.
These equations contain the Maxwell model as a special limit. We
discuss some results of plasticity which can be described by this
set of equations. We exploit the model equations for the simple
examples: straining of a slab and a rod. We find that necking
manifests always itself (not as a result of instability), except if
the very special constant-velocity stretching process is imposed.
\end{abstract}
\maketitle Plastic materials exhibit several features which are not
present in the usual liquids or solids. Their dynamics consist in a
nontrivial mixture of liquid-like and solid-like behaviors.
Understanding   plasticity in metal industry and, in general, in
technology, is of a paramount importance. Nonetheless, to date no
universal dynamical equations describing plastic materials like
Navier-Stokes equations for fluids, and Lam\'e equations for elastic
solids, are available. Under strain, plastic material may exhibit
elastic-like behaviors, yield stress, flowing behaviors, nonlinear
engineering strain-stress relation, and so on~\cite{Malverne}. A
major goal in material science is the description of these phenomena
in terms of dynamical evolution equations for relevant variables,
namely  the velocity, stress, and the analogues of strain.

There has been important contributions to the theory of plasticity,
especially for crystalline materials in terms of
dislocations~\cite{LL7,Kubin,Groma}. However, there is no need to
evoke dislocations (if ever this notion has a meaning) for amorphous
materials, and thus the question arises of how a corresponding
theory can be build at the continuum level. This question has known
recently an upsurge of interest~\cite{ELP}. An essential issue when
addressing the question of plasticity is the distinction between
crystalline solids and amorphous materials. Elastically deformed
monocrystals are in a metastable state. Their plastic flow takes
place only upon creation of dislocations, and is thus a nonlinear
process. Conversely, plastic flow of amorphous materials should
occur, in principle, linearly with respect to the applied stress. In
crystals, an additional field, namely dislocation density, is
introduced which couples to the elastic  as well as to plastic
distorsions~\cite{Kubin,Groma}. It is proposed here that one can
derive a plastic continuum theory for amorphous materials, without
evoking neither dislocation density, nor an internal variable that
is distinct from the variables describing usual kinematics of
plasticity. The evolution equations can be written in a closed form
in terms of the elastic and plastic distorsions only. The concept of
distortion~\cite{LL7} used here was introduced by Kr\"oener and
Rieder in~1956 \cite{Kroner}, and will constitute our basic
definition of the plastic flow variable.

If $u_k$ denotes the $k^{th}$ component of the displacement field,
then the total distorsion tensor reads~$\partial_iu_k$. For a purely
elastic solid, and in the small deformation regime, the symmetrized
part ${u_{ik}=(\partial_iu_k+\partial_ku_i)/2}$ is the strain
tensor. If plastic flow is involved, the total distorsion tensor
$\partial_i u_k$ is a sum of the plastic flow contribution
$w^{pl}_{ik},$ and the elastic part $w_{ik}$ (see~\cite{LL7})
\begin{equation} \label{d}
\partial_iu_k=w^{pl}_{ik}+w_{ik}.
\end{equation}
The symmetrical part of the elastic distortion defines the strain
tensor
\begin{equation}\label{uik}u_{ik}=\frac{1}{2}(w_{ik}+w_{ki}).\end{equation}
The constrain ${w^{pl}_{ii}=0}$ is usually supposed. Then from
Eqs.(\ref{d},\ref{uik}) it follows ${u_{ii}=\partial_iu_i.}$

Usual elasticity can be presented in a Lagrangian manner by
introducing an energy and a dissipative function. The elastic
energy of a solid is given by
\begin{equation} \label{U}
\frac{\lambda}{2}u^2_{ll}+\mu u^2_{ik},
\end{equation}
here $\lambda,\mu$ are Lam\'e coefficients, and the kinetic energy
is $\rho{\bf v}^2/2,$ where $\rho$ is the density of the material.
The dissipation function reads for a viscous solid
\begin{equation} \label{R0}
\eta^s\left(\dot{u}_{ik}-\frac{1}{3}\delta_{ik}\dot{u}_{ll}\right)^2
+\frac{\zeta}{2}\dot{u}_{ll}^2
\end{equation}
In the presence of a plastic flow one must introduce a new
additional dissipative part related to plasticity. The system is
described by three independent tensors, namely $u_{ik}$, and
$w^{pl\pm}_{ik}$ which are the symmetric $(+),$ and antisymmetric
$(-)$ parts of the plastic distortion tensor, respectively.   From
the basic kinematic relation~(\ref{d}) and the strain tensor
definition~(\ref{uik}) it follows
\begin{equation}\label{da}w^{pl+}_{ik}=
\frac{1}{2}(\partial_iu_k+\partial_ku_i)-u_{ik}.\end{equation} We
expect the dissipation to consist of a quadratic form of these
quantities, that we write as
\begin{equation} \label{R1}
2\alpha\dot{u}_{ik}\dot{w}^{pl+}_{ik}
+\eta\left(\dot{w}^{pl+}_{ik}\right)^2
+\gamma\left(\dot{w}^{pl-}_{ik}\right)^2\end{equation} This is the
dissipation corresponding to plastic flow. Stability criteria for
the dissipation function enforce ${\eta^s>0,}$ ${\zeta>0,}$
${\eta\eta^s>\alpha^2,}$ ${\gamma>0.}$ Note that there is only one
dilatational viscosity constant~$\zeta.$ All other constants are
related with shear motions. As we will see below in the liquid limit
the constant~$\eta$ is a usual hydrodynamic viscosity.

The strategy now consists in performing variations of the total
Lagrangian with respect to the independent variables. Variation with
respect to $\delta{\bf u},$ with ${\delta w^{pl}_{ik}=0},$ yields,
upon using, ${2\delta u_{ik}=\partial_i\delta u_k+\partial_k\delta
u_i,}$ the momentum conservation law
\begin{equation}\label{eleq}\rho\ddot{u}_i=\partial_k\sigma_{ik}
\end{equation} with the stress
tensor $\sigma_{ik}$ consisting of a sum of the usual elastic part
as well as the dissipative part with the usual solid viscosity
terms, and an additional plastic term
$$\sigma_{ik}=\lambda u_{ll}\delta_{ik} +2\mu u_{ik}+$$
\begin{equation}\label{sig} +2\eta^s\dot{u}_{ik}
+\left (\zeta-\frac{2}{3}\eta^s \right
)\delta_{ik}\dot{u}_{ll}+2\alpha\dot{w}^{pl+}_{ik}.
\end{equation}
Variation with respect to ${\delta w^{pl+}_{ik}},$ with
${\delta{\bf u}=0},$ and ${\delta w^{pl-}_{ik}=0}$ (in that case
${\delta u_{ik}=-\delta w^{pl+}_{ik}}$) provides us with
\begin{equation}\label{w-pl}\tilde{\sigma}_{ik}=
\alpha\left(2\dot{u}_{ik}-\frac{2}{3}\delta_{ik}\dot{u}_{ll}\right)
+2\eta\dot{w}^{pl+}_{ik},
\end{equation}
here $\tilde{\sigma}_{ik}$ is the traceless part of $\sigma_{ik}$.

Finally, variation with respect to ${\delta w^{pl-}_{ik}},$ with
${\delta{\bf u}=0},$ and ${\delta w^{pl+}_{ik}=0}$ (in that case
${\delta u_{ik}=0}$) leads to ${\dot{w}^{pl-}_{ik}=0}$ (a direct
consequence of the absence of dissipation for rigid rotation; note
also that energy does not depend on that mode).

An important remark is in order. Differentiating (\ref{d}) with
respect to time one obtains
\begin{equation}
\partial_k v_i+ \partial _ k v_i= - 2j_{ik}^{pl}+2\dot
u_{ik}\label{langer}
\end{equation}
where ${j_{ik}^{pl}\equiv-{\dot w^{pl+}_{ik}},}$ is the plastic
current. This equation (see also \cite{LL7}), apart from the
(conventional) minus sign in front of $j_{ik}^{pl},$ bears
resemblance with~Eq.(3) of Ref.~\cite{ELP}. There is, however, a
fundamental difference. Indeed, Eq.(3) of Ref.~\cite{ELP} uses the
kinematic condition (\ref{langer}), {\it plus}  Hooke's law, where
$u_{ik}$ is assumed to be related to the stress tensor by
$$u_{ik}={\tilde \sigma_{ik}\over 2\mu} -{p\over 2K},$$ where
${p=-\sigma_{kk}/2},$ is the pressure, and $K$ and $\mu$ are the
compressibility and the shear modulus (note that a 2D geometry is
assumed in Ref.~\cite{ELP}). In the present study we do not
postulate {\it a priori} a Hooke's relation, since both elastic and
plastic contributions are embedded together within the total
distortion tensor $\partial_i u_k.$  The relation between $u_{ik}$
and $\sigma_{ik}$ follows here as a consequence of the Lagrangian
formulation, and the relationship between these two quantities is
provided by (\ref{sig}) (showing that a measure of the stress is a
combination of elastic and plastic deformations).

It is possible to express the plastic distortion tensor in terms of
other quantities. From Eqs.(\ref{sig}-\ref{w-pl}) we may express
$\sigma_{ik}$ in terms $u_{ik}$ and its time derivative. It is
convenient to split the stress tensor $\sigma_{ik}$ into a traceless
and a pressure-like term (actually the trace of $\sigma_{ik}$):
\begin{equation}\label{sigm}\sigma_{ik}=
\frac{1}{3}\delta_{ik}\sigma_{ll}+\tilde{\sigma}_{ik}\end{equation}
 The trace has a usual elastic (including the dissipative part)
form
\begin{equation}\label{sigll}\sigma_{ll}=
(3\lambda+2\mu)u_{ll}+3\zeta\dot{u}_{ll}.\end{equation} The
traceless parts of the stress  tensor are connected with each other
and with spatial gradients of the velocity by the following two
relations
\begin{equation} \label{sigma1}
(\eta-\alpha)\tilde{\sigma}_{ik}=2\eta\mu\tilde{u}_{ik}+
2(\eta\eta^s-\alpha^2)\dot{\tilde{u}}_{ik},
\end{equation}
\begin{equation} \label{pv}
\tilde{\sigma}_{ik}+2(\eta-\alpha)\dot{\tilde{u}}_{ik}=\eta\left(
\partial_iv_k+\partial_kv_i-\frac{2\delta_{ik}}{3}\partial_lv_l\right).
\end{equation}
The first relation is obtained by  expressing
$\dot{\omega}^{pl}_{ik}$ from Eq.~(\ref{w-pl}) and inserting the
resulting relation  into (\ref{sig}). The second one follows from
Eq.~(\ref{w-pl}) by using relation~(\ref{da}). The  set of
Eqs.~(\ref{eleq},\ref{sigm}-\ref{pv}) defines space-time evolution
of displacement vector ${\bf u},$ strain $u_{ik}$ and the stress
tensor $\sigma_{ik}$. This constitutes a complete set of equations
for the three (vectorial and tensorial ) quantities ${\bf u},$
$u_{ik}$ and $\sigma_{ik}$ (we could, of course, alternatively use
other quantities like $w_{ik}^{pl+}$). Note that Eq.~(\ref{pv}) has
some similarity with the Maxwell model, used to describe plasticity
with a yield stress in some models~\cite{saramito}. There is an
important difference, however. Instead of the $\dot \sigma_{ik}$ on
the l.h.s. we have $\dot u_{ik}.$ Again this consistently follows
 from the Lagrangian formulation.

The Maxwell model of liquids with high viscosity can be obtained
from our equations only if one consider the incompressible limit
and set $\alpha=\eta^s=\zeta=0.$ This leads, from
Eq.~(\ref{sigma1}), to ${\sigma_{ik}=2\mu u_{ik}}.$ Eq.~(\ref{pv})
reduces to the well known Maxwell form
\begin{equation}\label{M}
\dot{\sigma}_{ik}+\mu\eta^{-1}\sigma_{ik} =
\mu(\partial_i\dot{u}_k+\partial_k\dot{u}_i).
\end{equation}

Let us present few examples where we could obtain an exact solution
of the plastic dynamics. Consider an induced oscillatory motion in
the material. We assume a semi-infinite medium bounded by a planar
surface which undergoes oscillations in its own $xz-$plane:
${u_x(t,0)=u\cos\omega t.}$ In this case the nonzero components of
the fields are $u_x,$ $u_{xy},$ $w_{xy},$ $w_{yx},$ $w^{pl+}_{xy},$
and $\sigma_{xy}.$ Then the set of
Eqs.~(\ref{eleq},\ref{sigma1},\ref{pv}) reads:
\begin{equation}\label{eleq1}\rho\ddot{u}_x=
\partial_y\sigma_{xy}=\partial_y\frac{2\eta\mu u_{xy}
+2(\eta\eta^s-\alpha^2)\dot{u}_{xy}}{\eta-\alpha};\end{equation}
\begin{equation}\label{sigma2}
\mu u_{xy}+(\eta-2\alpha+\eta^s)\dot{u}_{xy}
=\frac{\eta-\alpha}{2}\partial_yv_x.\end{equation} It is then found
 that each field is a linear combination of the complex modes
${\propto\exp(-i\omega t+i\kappa y)},$ where $\kappa$ is defined by
\begin{equation}\label{kappa}
\kappa^2=(\kappa'+i\kappa'')^2=i\frac{\omega\rho}{\eta}\left(1-
\frac{i\omega(\eta-\alpha)^2}{\eta\mu-i\omega(\eta\eta^s-\alpha^2)}
\right).\end{equation} For example, the displacement field is
${u_x=u\cos(\omega t - \kappa'y)\exp(-\kappa''y).}$ The low
frequency limit recovers a known Stokes result for a shear viscous
mode in liquids (see \S24 in~\cite{LL6}). Elastic solid behavior (an
emission of shear sound) corresponds to the limit of high plastic
viscosity ${\eta\rightarrow\infty}.$

\begin{figure}[hbt]
\includegraphics[width=\hsize]{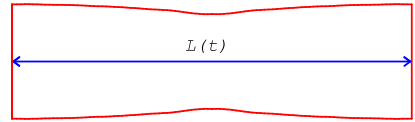}
\caption{A sketch of the geometry under
consideration.\label{stripe}}
\end{figure}

We would like to point out some  results that can be captured
analytically in some special limit. The long time behavior of a slab
under tension, is expected to be dominated by plastic flow.
Ultimately, the plastic flow should look-like a hydrodynamical flow.
Let us concentrate on this limit. Consider a plate (or a rod)  of a
plastic material with free surfaces (Fig.~\ref{stripe}). This is a
similar geometry to that treated in Ref.~\cite{ELP}. The plate is
stretched along the $x$ direction. For a flat geometry we have
obtained an exact solution with the plate thickness $h(t)$ that
depends only on the $t$ variable. This type of solution exists only
in the case where the stretching occurs at a given {\it constant}
velocity. Let us first motivate the solution on the basis of
symmetries. Because of the axial symmetry with respect to the $y$
axis at ${x=0},$ $v_x$ must be zero on that line. For constant $h$
there is a simple solution that fulfills that symmetry, ${v_x=cx},$
where $c(t)$ is for the moment an arbitrary function of time. From
incompressibility condition we have ${v_y=-cy+g(x,t),}$ where $g$ is
{\it a priori} an arbitrary function of $x$ and $t.$ Symmetry with
respect the middle line ${y=0},$ enforces ${g=0}.$

We straightforwardly obtain from the Navier-Stokes equation the
pressure field $$p=-\rho(\dot{c}+c^2)\frac{x^2}{2}+
\rho(\dot{c}-c^2)\frac{y^2}{2}+f(t),$$ where $f(t)$ is a function
of time to be determined below. At the free surface the normal
component of the stress (the tangential vanishes automatically)
must vanish. This is easily computed from the above result by
using the definition ${\sigma_{yy}=-p+2\eta\partial_yv_y=-p-2\eta
c.}$ Imposing ${\sigma_{yy}=0}$ on the free surfaces at ${y=\pm
h,}$ at any $x,$ we obtain ${\dot{c}+c^2=0.}$ This provides us
with ${c=(t-t_0)^{-1}},$ where $t_0$ is a constant of integration.
It is convenient to measure the time from the moment $t_0,$ so we
will set ${t_0=0}.$
For the length of the strip $L,$ one has ${\dot{L}=cL,}$ so that
${L=st,}$ with constant velocity $s.$ Due to mass conservation one
obtains that ${h=\gamma/t,}$ where  the total volume is defined as
$\gamma s.$ The solution  corresponds to stretching plastic flow for
${t>0}$, and ${s,\gamma>0,}$ while for ${t<0},$ and ${s,\gamma<0,}$
it represents a contracting flow.

Reporting this solution into the  above-mentioned boundary
conditions fixes the function ${f(t)=\rho\gamma^2t^{-4}-2\eta
t^{-1}}.$ The $xx$-component of the stress tensor is
${\sigma_{xx}=-p+2\eta\partial_xv_x=-p+2\eta c.}$ The total force
of stretching is defined as
\begin{equation}\label{F2}
F_{2d}\equiv\int_{-h}^h dy\sigma_{xx}= -{4\over
3}\rho{\gamma^2h\over t^4}+{8\eta h\over t}.
\end{equation}
The first part is inertial, while the second one is viscous. The
viscous part dominates at long time such that
${t\gg\gamma^{2/3}\nu^{-1/3},}$ where ${\nu=\eta/\rho}$ is the
kinematic viscosity.

It is a simple matter to extend the calculation to a cylindrical
geometry (stretching of a rod) following  precisely the same line.
We only give the results: The velocity is given by ${v_z=cz},$
${v_r=-cr/2},$ with ${c(t)=t^{-1}}$. The pressure reads
$$p=\frac{3\rho}{8t^2}\left(\frac{\gamma^2}{t}-r^2\right)-\frac{\eta}{t}.$$
We obtain for the length and the radius  ${L=st},$
${a=\gamma/\sqrt{t}.}$ The volume is given by ${\pi a^2L=\pi
s\gamma^2}.$ The total axial force in the rod stretching problem is
\begin{equation}\label{F3}
F_{3d}\equiv 2\pi\int _{0}^a rdr \sigma_{zz}=
-\frac{3}{16}\frac{\rho\gamma^2}{t^3}S+3\frac{\eta}{t}S,
\end{equation}
where ${S=\pi a^2}$ is a rod cross section area.

The above solutions exists only for a constant velocity stretching.
The question thus naturally
arises of what happens if an other process is imposed. This is what
we would like to investigate now.
Following Ref.\cite{ELP}, if the lateral boundary of the plate moves
at a pre-determined strain rate $\dot L/L=\Omega=const.,$ then our
result shows that a  homogeneous thinning of the strip  is not
possible. Thus  a modulated strip prevails. This is a precursor of
the necking problem. Thus  necking appears here as natural
phenomenon due to material flowing \cite{Hutchinson} whenever the
stretching is not performed at a constant speed. The necking is not
related with an instability~\cite{ELP}, but rather the fact that a
homogeneous thinning does not exist (except if ${L(t)\propto t}$) in
plastic dynamics.

Let us investigate the stability of the homogeneous solutions. We
consider deviations with large wavelengths as compared with the
layer thickness. By analogy with the theory of shallow water (see
\cite{LL6} \S108) we derive effective hydrodynamic equations in
terms of the thickness $h,$ and velocity $v$ along $x$. Mass
conservation yields
\begin{equation}\label{mass}
\dot{h}+\partial_x(hv)=0.
\end{equation}
Note, that when the viscous term dominates, one can rewrite the
total force (\ref{F2}) as follows ${F_{2d}=-8\eta\partial_th}.$ In a
general case of inhomogeneous shape one should have
${F_{2d}=-8\eta(\dot{h}+v\partial_xh)}.$ Consequently the momentum
conservation law in this purely viscous limit has the form
\begin{equation}\label{2dNS}
h(\dot{v}+v\partial_xv)=-8\nu\partial_x(\dot{h}+v\partial_xh).
\end{equation}

From Eqs.(\ref{mass},\ref{2dNS}) one obtains upon linearization
about the 2D homogeneous solution:
${t\delta\dot{h}+x\partial_x\delta h+\gamma\partial_x\delta v+\delta
h=0}$, and ${\gamma(t\delta\dot{v}+x\partial_x\delta v+\delta v)=
-8\nu t\partial_x(t\delta\dot{h}+ x\partial_x\delta{h}).}$
Introducing the new coordinate ${\tilde{x}=x/t}$ one arrives at:
$t^2\delta\dot{h}+\gamma\delta v'+t\delta h=0,$ and
$\gamma(t\delta\dot{v}+\delta v)=-8\nu t\delta\dot{h}'.$ Taking the
Fourier transform ${\propto\exp(iq\tilde{x})},$ we find for a given
mode with wavenumber  $q$ of one of the fields, say $\delta h:$
${t^2\delta \ddot{h}+4(t+\tau)\delta\dot{h}+2\delta h=0},$ here
${\tau=2\nu q^2.}$  This equation has a first integral
${t^2\delta\dot{h}+2t\delta h+4\tau\delta h=C,}$ where $C$ is a
constant. One obtains finally  $$\delta h
=\frac{C}{t^2}\exp\left(\frac{4\tau}{t}\right)\int^t_{t^*}
\exp\left(-\frac{4\tau}{t}\right)dt,$$ where $t^*$ is a second
constant of integration. At large time, ${t\gg\tau},$ the deviation
amplitude is ${\delta h=Ct^{-1}\{1-4\tau t^{-1}\ln(t/et^*)\}.}$ It
decays mostly as $t^{-1}$ like the strip thickness $h.$
Consequently, one may say that the solution is marginally stable for
large wavelength fluctuations.

The 3D problem of the rod stretching with small inhomogeneity can
be formulated in the  same manner (Eqs.(\ref{mass},\ref{2dNS})).
The results are identical to the 2D ones, if one makes the
substitutions ${h\rightarrow S},$ and ${8\nu\rightarrow3\nu}.$

We have solved Eqs.(\ref{mass})-(\ref{2dNS}) numerically in the
fully nonlinear regime. We give here the major results: (i) If one
imposes a constant stretching we find that the ultimate stage is a
homogeneous thickness that decreases in time as $1/t,$ in agreement
with our analytical results. (ii) Starting from a small perturbation
(of sine type), we observe marginal stability. (iii) Most
importantly, we have found that if the stretching velocity is not
constant the ultimate stage is a modulated thickness, of necking
type. In fact, Fig.\ref{stripe} represents the result of our
numerical solution, that exhibits necking in the case of initially
flat plate and ${\dot L/L=const.}$ (same stretching law as in
Ref.~\cite{ELP}). This behavior is found for various initial
conditions, and  (non constant) stretching laws. Thus necking seems
to be a robust feature, which takes place whenever the stretching is
non constant.

It should be mentioned that here our plastic equations  have been
written by disregarding the so-called objective derivative (we have
used ordinary derivatives) for tensors. One alterative in order to
confer an objective form to these equations, is to replace the time
derivative of tensors by the so-called co-rotational
derivative\cite{Bird1}, as is done in \cite{ELP}. We shall report on
full numerics of our completed set of equations
(\ref{eleq},\ref{sigm}-\ref{pv}) in the future.

We acknowledge CNRS, Univ.  J. Fourier, and CNES for financial
support.

\end{document}